\documentclass[11pt,canadian,preprint,preprintnumbers,amsmath,amssymb,nofootinbib,superscriptaddress]{revtex4}
\usepackage[latin9]{inputenc}
\usepackage[a4paper]{geometry}
\usepackage{color}
\usepackage{babel}
\usepackage{rotating}
\usepackage{amsmath}
\usepackage{amssymb}
\usepackage{graphicx,subfigure}
\usepackage{esint}
\usepackage{epstopdf}
\usepackage[unicode=true,pdfusetitle,
 bookmarks=true,bookmarksnumbered=false,bookmarksopen=false,
 breaklinks=false,pdfborder={0 0 1},backref=false,colorlinks=false]
 {hyperref}

\makeatletter
\@ifundefined{textcolor}{}
{%
 \definecolor{BLACK}{gray}{0}
 \definecolor{WHITE}{gray}{1}
 \definecolor{RED}{rgb}{1,0,0}
 \definecolor{GREEN}{rgb}{0,1,0}
 \definecolor{BLUE}{rgb}{0,0,1}
 \definecolor{CYAN}{cmyk}{1,0,0,0}
 \definecolor{MAGENTA}{cmyk}{0,1,0,0}
 \definecolor{YELLOW}{cmyk}{0,0,1,0}
}

\usepackage{color}\usepackage{graphics}\usepackage{dcolumn}\usepackage{bm}\usepackage{afterpage}

\numberwithin{equation}{section}

\makeatother

\begin{document}

\allowdisplaybreaks

\title{Numerical Boson Stars with a Single Killing Vector I: the $D\ge5$ Case}

\author{Sean Stotyn}

\email{sstotyn@phas.ubc.ca}

\affiliation{Department of Physics and Astronomy, University of British Columbia,\\
 Vancouver, British Columbia, Canada, V6T 1Z1}

\author{C. Danielle Leonard}

\email{danielle.leonard@astro.ox.ac.uk}

\affiliation{Department of Physics, University of Oxford, Oxford, United Kingdom,
OX1 3RH}

\author{Marius Oltean}

\email{moltean@physics.mcgill.ca}

\affiliation{Department of Physics, McGill University, Montreal, Quebec, Canada,
H3A 2T8}

\author{Laura J. Henderson}
\email{l7hender@uwaterloo.ca}

\author{Robert B. Mann}
\email{rbmann@sciborg.uwaterloo.ca}

\affiliation{Department of Physics and Astronomy, University of Waterloo,\\
 Waterloo, Ontario, Canada, N2L 3G1}

\date{\today}

\begin{abstract}
We numerically construct asymptotically anti-de Sitter boson star solutions using a minimally coupled $\frac{D-1}{2}$-tuplet complex scalar field in $D=5,7,9,11$ dimensions. The metric admits multiple Killing vector fields in general, however the scalar fields are only invariant under a particular combination, leading to such boson star solutions possessing just a single helical Killing symmetry. These boson stars form a one parameter family of solutions, which can be parametrized by the energy density at their center.  As the central energy density tends to infinity, the angular velocity, mass, and angular momentum of the boson star exhibit damped harmonic oscillations about finite central values, while the Kretschmann invariant diverges, signaling the formation of a black hole in this limit.
\end{abstract}

\maketitle

\newpage{}

\section{Introduction}

Boson stars are smooth, horizonless geometries composed of self-interacting and self-gravitating bosonic matter\cite{Kaup:1968zz,Ruffini:1969qy}.  They describe localized bundles of field energy that typically do not display a sharp edge, like those of ordinary stars or neutron stars.  It is currently unknown whether boson stars are physically realized in our universe, but they have nevertheless maintained a steady level of interest both from a theoretical gravitational point of view and from an astrophysical point of view.  In the latter context, boson stars in four dimensional asymptotically flat space-time have been put forth as candidates for dark matter halos that may help explain galaxy rotation curves -- see Ref \cite{Jetzer:1991jr} for a review.  While boson stars in an excited state typically produce a more physically realistic, flatter rotation curve than boson stars in the ground state, such excited states are known to decay to the ground state unless the boson stars are in rather particular mixed states \cite{Bernal:2009zy}.  If not coupled to a Maxwell field, they also provide dark alternatives to astrophysical black hole candidates, which could potentially be discerned by gravitational wave astronomy \cite{Berti:2006qt,Kesden:2004qx}.  On the purely theoretical side, boson stars in asymptotically anti-de Sitter (AdS) space-times are believed to play important roles in holographic gauge theories through the AdS/CFT correspondence.  There is much work yet to be done in this respect, but some progress has been made, at least in mapping out the possible boson star solutions.  For a detailed review of boson stars and their various applications, we refer the reader to Ref \cite{Liebling:2012fv}.

There are a wide range of boson star solutions, which can be composed of a complex scalar field with mass\cite{Astefanesei:2003qy,Brihaye:2013hx} or without\cite{Stotyn:2011ns,Dias:2011at}, with self-interactions\cite{Colpi:1986ye,Ho:2002vz}, with gauge charges\cite{Brihaye:2004nd,Dias:2011tj}, with rotation\cite{Stotyn:2011ns,Dias:2011at}, with de Sitter\cite{Fodor:2010hg,Cai:1997ij}, flat\cite{Colpi:1986ye,Hartmann:2012gw}, or anti-de Sitter\cite{Astefanesei:2003qy,Bjoraker:1999yd} boundary conditions, etc.  Despite this diversity, boson stars all have one important feature in common: since they are horizonless, they are zero temperature objects that describe finite energy excitations above the vacuum state.  The stability of these objects is of central interest if AdS boundary conditions are present.  If they are perturbatively unstable to the formation of a black hole, then they are themselves of little interest gravitationally.  If they are perturbatively stable then their non-linear stability determines whether the corresponding state in the holographic dual CFT thermalizes and on what time-scale.  Due to the evidence of a gravitational turbulent instability in asymptotically AdS space-times \cite{Dias:2011ss,Bizon:2011gg,Jalmuzna:2011qw,Maliborski:2013jca}, one might expect all AdS boson stars to be non-linearly unstable to black hole formation.  However, recently in \cite{Buchel:2013uba} a wide range of initial data were discovered such that boson stars are immune to the turbulent instability, leading on the gauge theory side to a family of strongly coupled CFT states that do not thermalize in finite time.  Furthermore, the analysis of \cite{Dias:2012tq} suggests that the turbulent instability of global AdS is due to the high level of symmetry: the normal mode frequencies are all integer multiples of the AdS frequency, leading to a large number of resonances responsible for the nonlinear instability.  Solutions possessing less symmetry, such as boson stars, are not plagued by this problem and tend to be nonlinearly stable as a result.  Although a clear mapping between boson star properties and physically observed states in CFTs is currently unknown, it is hoped that a better understanding of the gravitational aspects of boson stars will lead to insights into physically realizable systems.

Most of the boson star solutions constructed to date have relied on a relatively high level of symmetry to yield equations that can feasibly be solved numerically.  Indeed, space-time symmetries often play an indispensable role in constructing analytic solutions to the Einstein equations. Such symmetries readily manifest in the familiar form of Killing vector fields, which generate the space-time's isometry group.  Furthermore, various theorems show that Killing symmetries are ubiquitous under physically reasonable assumptions.  For instance, the rigidity theorem states that if a space-time is stationary, it must also possess an axis of symmetry, leading to a minimum of 2 Killing symmetries\cite{Hawking:1971vc,Hollands:2006rj,Moncrief:2008mr}.  Concordantly, it has become an almost universal feature of exact solutions to Einstein gravity to admit multiple Killing fields.

However, this need not necessarily be the case: for example, the gravitational field outside of a generic matter source may only have approximate Killing symmetries, or none at all.  Even in more idealized settings, there is considerable interest in exact solutions which possess only a single Killing field.  For instance, due to a judicious choice of metric and scalar field ansatz first put forth in \cite{Hartmann:2010pm}, spinning boson star solutions exhibiting a single helical Killing vector have been found in 5 dimensions with anti-de Sitter (AdS) boundary conditions \cite{Dias:2011at}; the scalar field has a harmonic time dependence that breaks the continuous rotational symmetry.  In \cite{Stotyn:2011ns} these techniques were used to construct analogous boson stars in odd dimensions $D\ge5$ in the perturbative regime where energy and angular momentum are low.  However full numerical results away from the perturbative regime have yet only been presented in the 5 dimensional treatment in \cite{Dias:2011at}.  
In this paper, we complete this analysis and obtain numerical solutions for asymptotically AdS boson stars with a single Killing field in $D=7,9,11$ dimensions; this one parameter family of solutions is parameterized by the central energy density of the boson star.  This central energy density is unbounded above and in the limit that it tends to infinity, the mass, angular momentum, and angular velocity all tend to finite values, while the Kretschmann scalar at the centre diverges.  Analogous boson stars in $D=3$ have qualitatively distinct features and require separate considerations, even in the perturbative regime \cite{Stotyn:2012ap}; the numerical construction of these lower dimensional boson star solutions is discussed in a separate paper \cite{Stotyn:2013spa}.  We focus on odd-dimensions since there is a useful ansatz for the scalar fields (given in section 2) that allows the full solutions to have only a single Killing vector, but yields a stress-energy with the same symmetries as the metric.

The remainder of this paper is structured as follows: in section \ref{Setup} we present our ansatz and extract the ordinary differential equations (ODEs) and constraint equations we must solve. In section \ref{BosonConditions}, we discuss the boundary conditions and the physical properties of our boson stars. In section \ref{numconst} we describe our numerical methods. In section \ref{Results} we present and discuss our results, and in section \ref{Conclusion} we provide some concluding remarks.

\section{Setup}

\label{Setup}

We begin with $D=n+2$ dimensional Einstein gravity with negative
cosmological constant minimally coupled to an $\frac{n+1}{2}$-tuplet
complex scalar field 
\begin{equation}
S=\frac{1}{16\pi}\int{d^{D}x\sqrt{-g}\left(R+\frac{n(n+1)}{\ell^{2}}-2\big|\nabla\vec{\Pi}\big|^{2}\right)}\label{eq:action}
\end{equation}
where we take the usual convention $\Lambda=-\frac{n(n+1)}{2\ell^{2}}$ with AdS length $\ell$ and restrict attention to $n=3,5,7,9$.
The equations of motion resulting from this action are $G_{ab}-\frac{n(n+1)}{2\ell^{2}}g_{ab}=T_{ab}$
and $\nabla^{2}\vec{\Pi}=0$, where the stress tensor of the scalar
field is given by 
\begin{equation}
T_{ab}=\left(\partial_{a}\vec{\Pi}^{*}\partial_{b}\vec{\Pi}+\partial_{a}\vec{\Pi}\partial_{b}\vec{\Pi}^{*}\right)-g_{ab}\left(\partial_{c}\vec{\Pi}\partial^{c}\vec{\Pi}^{*}\right).\label{eq:Tab}
\end{equation}

We take the following ans\"atze for the metric and scalar
field 
\begin{equation}
ds^{2}=-f(r)g(r)dt^{2}+\frac{dr^{2}}{f(r)}+r^{2}\bigg(h(r)\big(d\chi+A_{i}dx^{i}-\Omega(r)dt\big)^{2}+g_{ij}dx^{i}dx^{j}\bigg)\label{eq:metric}
\end{equation}
\begin{equation}
\Pi_{i}=\Pi(r)e^{-i\omega t}z_{i},\quad\quad\quad\quad i=1...\frac{n+1}{2}\label{eq:ScalarField}
\end{equation}
where $z_{i}$ are complex coordinates such that ${\displaystyle \sum_{i}dz_{i}d\bar{z}_{i}}$
is the metric of a unit $n-$sphere. An explicit and convenient choice
for the $z_{i}$ is 
\begin{equation}
z_{i}=\left\{ {\genfrac{.}{.}{0pt}{0}{e^{i(\chi+\phi_{i})}\cos\theta_{i}{\displaystyle \prod_{j<i}\sin\theta_{j},\quad\quad\quad i=1...\frac{n-1}{2}}}{{e^{i\chi}{\displaystyle \prod_{j=1}^{\frac{n-1}{2}}\sin\theta_{j}}},\quad\quad\quad\quad\quad i=\frac{n+1}{2}}}\right.\label{eq:zi}
\end{equation}
in which case ${\displaystyle \sum_{i}dz_{i}d\bar{z}_{i}=(d\chi+A_{i}dx^{i})^{2}+g_{ij}dx^{i}dx^{j}}$
is the Hopf fibration of the unit $n-$sphere where 
\begin{equation}
A_{i}dx^{i}=\sum_{i=1}^{\frac{n-1}{2}}{\cos^{2}\theta_{i}\left[\prod_{j<i}\sin^{2}\theta_{j}\right]d\phi_{i}}
\end{equation}
and $g_{ij}$ is the metric on a unit complex projective space $\mathbb{CP}^{\frac{n-1}{2}}$.  In these coordinates, $\chi$ and the $\phi_{i}$ all have period $2\pi$ while the $\theta_{i}$ take value in the range $[0,\frac{\pi}{2}]$.

The form of the scalar fields is crucial to this construction and
was first considered in \cite{Hartmann:2010pm}: it is clear from
Eq. (\ref{eq:ScalarField}) that the scalar fields can be viewed as
coordinates on ${\mathbb{C}}^{\frac{n+1}{2}}$.  For each value of $r$, $\vec{\Pi}$ traces
out a round $n$-sphere with a time-varying but otherwise constant
phase. On the other hand, constant $r$ surfaces in the metric (\ref{eq:metric})
correspond to squashed rotating $n$-spheres. The stress tensor has
the same symmetries as the metric (\ref{eq:metric}) since the first
term is the pull-back of the round metric of the $n$-sphere and the
second term is proportional to $g_{ab}$.

Although the matter stress tensor has the same symmetries as the metric,
the scalar fields themselves do not. Indeed, the metric (\ref{eq:metric})
is invariant under $\partial_{t}$, $\partial_{\chi}$ as well as
the rotations of $\mathbb{CP}^{\frac{n-1}{2}}$ while the scalar field
(\ref{eq:ScalarField}) is only invariant under the combination 
\begin{equation}
K=\partial_{t}+\omega\partial_{\chi}.\label{eq:KV}
\end{equation}
Therefore, any solution with non-trivial scalar field will only be
invariant under the single Killing vector field given by (\ref{eq:KV}).

The equations of motion yield the following system of coupled second
order ODEs:
\begin{equation}
\begin{split}f''-\frac{6f'}{fr}\left(\frac{rf'}{6}-\frac{f}{6}+\Xi\right)+\frac{4h'}{r}+\frac{n^{2}-1}{r^{2}}+\frac{n^{2}-1}{\ell^{2}}+\frac{8\Pi'\Pi}{r}+\frac{4\Pi^{2}(\omega-\Omega)^{2}}{fg}\\
-\frac{8\Xi^{2}}{fr^{2}}-\frac{4\Pi^{2}\left(1+\frac{(n-1)}{2}h\right)}{hr^{2}}-\frac{2(n-3)\Xi}{r^{2}}=0, & \label{eq:fEq}
\end{split}
\end{equation}
\begin{equation}
\begin{split}g''-g'\left(\frac{4\Xi}{fr}+\frac{g'}{g}-\frac{1}{r}\right)-4g\left(\frac{\big(\Xi r^{\frac{n-1}{2}}\sqrt{h}\big)'}{fr^{\frac{n+1}{2}}\sqrt{h}}+\frac{\frac{(n-1)}{2}h^{2}-\Pi^{2}}{fhr^{2}}+\frac{3(n+1)}{2f\ell^{2}}-\frac{(n-3)\Xi}{2fr^{2}}\right)\\
-\frac{8\Pi^{2}(\omega-\Omega)^{2}}{f^{2}}-\frac{hr^{2}\Omega'^{2}}{f}=0, & \label{eq:gEq}
\end{split}
\end{equation}
\begin{equation}
h''+\frac{h'}{r}-\frac{2h'}{fr}\left(\Xi+\frac{frh'}{2h}\right)+\frac{h^{2}r^{2}\Omega'^{2}}{fg}+\frac{4(1-h)}{fr^{2}}\left(\Pi^{2}+\frac{(n+1)}{2}h\right)=0,\label{eq:hEq}
\end{equation}
\begin{equation}
\Omega''+\frac{4\Pi^{2}}{fhr^{2}}(\omega-\Omega)+\Omega'\left(\frac{f'}{f}+\frac{2h'}{h}+\frac{2\Xi}{fr}+\frac{2n+1}{r}\right)=0,\label{eq:OmegaEq}
\end{equation}
\begin{equation}
\Pi''-\frac{2\Pi'}{fr}\left(\Xi-\frac{f}{2}\right)+\frac{\Pi(\omega-\Omega)^{2}}{f^{2}g}-\frac{\big(1+(n-1)h\big)\Pi}{fhr^{2}}=0,\label{eq:PiEq}
\end{equation}
where $\Xi=h+\Pi^{2}-\frac{n+1}{2}-\frac{(n+1)r^{2}}{2\ell^{2}}$
and a $'$ denotes differentiation with respect to $r$. 

In addition to these second order ODEs, the
Einstein equations further impose two first order ODEs in the form
of constraint equations, $C_{1}=0$ and $C_{2}=0$. Explicitly, these
are 
\begin{equation}
C_{1}=\frac{(f^{2}ghr^{2(n-1)})'}{fr^{2n-3}}+4gh\Xi, \label{eq:C1Eq}
\end{equation}
\begin{equation}
C_{2}=\frac{\Pi^{2}(\omega-\Omega)^{2}}{f^{2}g}+\Pi'^{2}-\frac{r^{2}h\Omega'^{2}}{4fg}+\frac{\Xi(r^{n+1}h)'}{fhr^{n+2}}+\frac{(hf)'h'}{4fh^{2}}+\frac{n(fhr^{n-1})'}{2fhr^{n}}+\frac{\frac{(n-1)}{2}h^{2}-\Pi^{2}}{fhr^{2}}+\frac{n+1}{2f\ell^{2}}. \label{eq:C2Eq}
\end{equation}
Furthermore, under the flow of the equations of motion above, these constraint equations obey
\begin{align}
C_1'={}&{\mathcal F}_1(r) C_1, \label{eq:C1deriv} \\
C_2'={}&{\mathcal F}_2(r)C_1+{\mathcal F}_3(r)C_2, \label{eq:C2deriv}
\end{align}
for some functions ${\mathcal F}_1,~{\mathcal F}_2,~{\mathcal F}_3$ that depend on the dimensionality of the space-time.  The explicit form of these functions is neither important nor illuminating, but what is important is that the derivatives of the constraints are proportional to the constraints themselves.

\section{Boundary Conditions and Physical Charges}

\label{BosonConditions}

In order to solve the equations of motion, we need to impose boundary conditions at the boson star origin, as well as asymptotically.  In this section we present these boundary conditions and we write the asymptotic charges in terms of them.

\subsection{Boundary Conditions at the Origin}

The geometry must be smooth and horizonless, which means that all
metric functions must be regular at the origin. Furthermore, due to
the slow physical rotation of points as $r\rightarrow0$, surfaces
of constant $t$ in the vicinity of the origin ought to be described
by round $n$-spheres with $r$ being the proper radial distance.
To find the boundary condition on $\Pi$, we multiply (\ref{eq:PiEq})
by $r^{2}$ and note that $\Pi$ must vanish at the origin in order
to yield consistent equations of motion. Thus, the boundary conditions at the origin take the form 
\begin{equation}
\begin{split}
\left.f\right|_{r\rightarrow0}=1+\mathcal{O}(r^{2}),\quad\left.g\right|_{r\rightarrow0}=g(0)+\mathcal{O}(r),\quad\left.h\right|_{r\rightarrow0}=1+\mathcal{O}(r^{2}),\label{eq:OriginBC}\\
\left.\Omega\right|_{r\rightarrow0}=\Omega(0)+\mathcal{O}(r),\quad\left.\Pi\right|_{r\rightarrow0}=q_{0}\frac{r}{\ell}+{\cal O}(r^{2}),\quad\quad\quad\quad
\end{split}
\end{equation}
for all $n$, where $q_{0}$ is a dimensionless parameter
such that the energy density of the scalar field, $T^{00}$, at the
origin is proportional to $q_{0}^{2}$. In fact, $q_{0}$ uniquely
parameterizes the one-parameter family of boson star solutions in
each dimension.  Formally, it is defined by $q_0\equiv\ell\Pi'(0)$.

\subsection{Asymptotic Boundary Conditions}

In order to simplify the asymptotic boundary conditions, we first
make note of a residual gauge freedom. It is straightforward to show
that the transformation 
\begin{equation}
\chi=\tilde{\chi}+\lambda t,\quad\quad\Omega(r)=\tilde{\Omega}(r)+\lambda,\quad\quad\omega=\tilde{\omega}+\lambda\label{eq:PsiGaugeFreedom}
\end{equation}
for some arbitrary constant $\lambda$, leaves both the metric (\ref{eq:metric})
and scalar field (\ref{eq:ScalarField}) unchanged. We find it convenient
in our numerical analysis to set $\lambda=\omega$ so that we can
set $\tilde{\omega}=0$: in this frame, the coordinates are rigidly
rotating asymptotically so that $\tilde{\Omega}(r)\rightarrow-\omega$
as $r\rightarrow\infty$. In what follows, we use $\tilde{\chi}$, $\tilde{\Omega}(r )$, and $\omega$ but we drop the tildes for notational convenience.

In the $r\rightarrow\infty$ limit the boundary conditions for the
boson star will asymptote to AdS with corrections for mass and angular
momentum, which determine the metric functions up to constants $C_{f},~C_{h},~\mathrm{and}~C_{\Omega}$.
The boundary condition on the scalar field is set by requiring $\Pi$
to be normalizable, which means it must decay like $r^{-(n+1)}$.
Explicitly, the asymptotic boundary conditions are given by 
\begin{equation}
\begin{split}
\left.f\right|_{r\rightarrow\infty}={} & \frac{r^{2}}{\ell^{2}}+1+\frac{C_{f}\ell^{n-1}}{r^{n-1}}+\mathcal{O}(r^{-n}),\quad\>\>\left.g\right|_{r\rightarrow\infty}=1-\frac{C_{h}\ell^{n+1}}{r^{n+1}}+\mathcal{O}(r^{-(n+2)}), \\
\left.h\right|_{r\rightarrow\infty}={} & 1+\frac{C_{h}\ell^{n+1}}{r^{n+1}}+\mathcal{O}(r^{-(n+2)}),\quad\quad\left.\Omega\right|_{r\rightarrow\infty}=-\omega+\frac{C_{\Omega}\ell^{n}}{r^{n+1}}+\mathcal{O}(r^{-(n+2)}),\label{eq:AsymptoticBC}\\
\left.\Pi\right|_{r\rightarrow\infty}={} & \frac{\epsilon\ell^{n+1}}{r^{n+1}}+\mathcal{O}(r^{-(n+2)}), 
\end{split}
\end{equation}
where $\epsilon$ is a dimensionless measure
of the amplitude of the scalar field at infinity. A perturbative
analysis of this problem \cite{Stotyn:2011ns} indicates that $\epsilon$ uniquely parameterizes the boson star solutions when the energy and angular momentum are low; however in the non-perturbative regime it does not, as will be explicitly demonstrated in section \ref{Results}.

\subsection{Physical Charges}

As we have seen, these boson stars are invariant under the single
Killing field (\ref{eq:KV}), that is, a linear combination of $\partial_t$ and $\partial_\chi$. However, the scalar field vanishes at infinity with sufficient fall-off to imply that $\partial_{t}$ and $\partial_{\chi}$
are each asymptotic Killing fields (the metric alone being invariant
under them). Thus they define conserved charges, which can be computed
via the Astekhar-Das formalism \cite{Ashtekar:1999jx,Das:2000cu}; in particular, $\partial_{t}$ and $\partial_{\chi}$ are readily associated with a conserved energy and angular momentum, respectively.  For details on how these charges are defined and computed, we refer the reader to Ref \cite{Stotyn:2011ns} and here simply state the result:
\begin{align} \label{EJ}
M & =\frac{(n+1)\pi^{\frac{n-1}{2}}\ell^{n-1}}{16\left(\frac{n+1}{2}\right)!}\big((n+1)C_{h}-nC_{f}\big),\\
J & =\frac{(n+1)^{2}\pi^{\frac{n-1}{2}}\ell^{n}C_{\Omega}}{16\left(\frac{n+1}{2}\right)!}.
\end{align}
Here, $C_{f},~C_{h},~\mathrm{and}~C_{\Omega}$ are the constants appearing
in the asymptotic boundary conditions (\ref{eq:AsymptoticBC}).

The existence of these asymptotic Killing symmetries also guarantees that the boson stars satisfy the first law of thermodynamics.  They have vanishing temperature and entropy, so the first law takes the form
\begin{equation}
dM=\omega dJ.
\end{equation}
The fact that these boson stars satisfy the first law is an important numerical tool.  Indeed one of the primary cross-checks on the validity of our numerical methods discussed below is an explicit verification that the first law holds to at least one part in $10^{6}$.

\section{Numerical Construction}

\label{numconst}

To numerically construct boson star solutions, we used a relaxation method on a Chebyshev grid with the metric and scalar field functions approximated by Chebyshev polynomials.  A detailed review of approximating analytic functions by Chebyshev polynomials as well as the Chebyshev relaxation procedure can be found in \cite{Pfeiffer:2005zm}.

In the case under consideration here there are 5 functions $\{f(r),g(r),h(r),\Omega(r),\Pi(r)\}$ defined over the domain $r\in[0,\infty)$.
To employ the Chebyshev relaxation procedure, we need to compactify the
domain, which we do by introducing the new coordinate, $y$, defined
by
\begin{equation}
y=\frac{r^{2}}{r^{2}+\ell^{2}}.
\end{equation}
Note that since $r$ takes value in the range $[0,\infty)$, $y$ takes value in the range $[0,1]$.  In terms of the Chebyshev grid points $x_k=\cos\left(\frac{k\pi}{N}\right)$, $k=0,...,N$, defined as the extrema of the $N^{\mathrm{th}}$ Chebyshev polynomial $T_N(x)$ over the range $[-1,1]$, we define our grid by the simple transformation $y_{k}\equiv\frac{x_{k}+1}{2}$.  

To obtain a set of analytic functions, we first extract the singular
behavior from each and introduce auxiliary functions with the boundary
conditions (\ref{eq:OriginBC}) and (\ref{eq:AsymptoticBC}) in mind. In terms of the coordinate $y$, this leads to
\begin{align}
 & f(y)=\frac{y}{1-y}+1+(1-y)^{\frac{n-1}{2}}q_{f}(y), \label{fq}\\
 & g(y)=1+(1-y)^{\frac{n+1}{2}}q_{g}(y), \label{gq}\\
 & h(y)=1+(1-y)^{\frac{n+1}{2}}q_{h}(y), \label{hq}\\
 & \Omega(y)=\frac{q_{\Omega}(y)}{\ell}, \label{Oq}\\
 & \Pi(y)=\sqrt{y}(1-y)^{\frac{n+1}{2}}q_{\Pi}(y). \label{Pq}
\end{align}
The set of functions $\{q_{f},q_{g},q_{h},q_{\Omega},q_{\Pi}\}$ are
analytic over the range $y\in[0,1]$; in the remaining discussion
the coordinate $y$ will be used, unless otherwise stated. In particular,
a prime $'$ will denote a derivative with respect to $y$.

To find the boundary conditions on the $q$ functions, we Taylor expand
the equations of motion (\ref{eq:fEq})--(\ref{eq:OmegaEq}) and the constraint equations (\ref{eq:C1Eq}) and (\ref{eq:C2Eq}) around the two boundary points, $y_0=0,1$ and require them to vanish order by order in $(y-y_0)$. This leads to the following nontrivial relationships between the various functions and their first derivatives: at $y_0=0$ 
\begin{align} \label{eq:BC1}
& q_{f}(0)=0,\qquad\qquad q_{g}'(0)-\frac{n+1}{2}q_{g}(0)-\frac{3}{n}q_{h}'(0)\big(1+q_{g}(0)\big)-\frac{2}{n}\big(1+q_{g}(0)\big)q_{0}^{2}=0, \\
& q_{h}(0)=0,\qquad\qquad q_{\Omega}'(0)-\frac{2}{n+3}q_{0}^{2}q_{\Omega}(0)=0,\qquad\qquad q_{\Pi}(0)-q_{0}=0,
\end{align}
and at $y_0=1$ 
\begin{align}
&q_{f}'(1)+\frac{n-1}{2}q_{f}(1)+q_{h}(1)=0,\qquad\qquad q_{g}(1)+q_{h}(1)=0, \qquad\qquad q_{\Omega}'(1)=0,\\
&q_{h}'(1)+\frac{n+1}{2}q_{h}(1)=0, \qquad\qquad q_{\Pi}'(1)+\frac{1}{2(n+3)}\big((n+2)^{2}-q_{\Omega}(1)^{2}\big)q_{\Pi}(1)=0. \label{eq:BC2}
\end{align}

Note that since the constraint equations (\ref{eq:C1Eq}) and (\ref{eq:C2Eq}) are explicitly satisfied at the boundary points by these boundary conditions, the derivatives of the constraints, (\ref{eq:C1deriv}) and (\ref{eq:C2deriv}),  are also satisfied, meaning that the constraints are guaranteed to be satisfied over the whole domain.

To solve the equations of motion, we approximated each of the analytic $q$-functions by an order $N$ expansion in terms of Chebyshev polynomials, as described in \cite{Pfeiffer:2005zm}.  Next, this expansion was inserted into the equations of motion at each of the $N-1$ interior Chebyshev grid points (excluding the boundary points $y=0,1$).  At the two boundary points, the Chebyshev expansions were inserted into the boundary conditions  (\ref{eq:BC1})--(\ref{eq:BC2}). The equations of motion and boundary conditions then became a system of non-linear algebraic equations for the spectral coefficients of the Chebyshev approximations. A Newton-Raphson method was then employed by linearizing the equations of motion with respect to the spectral coefficients.

The initial seed solution at $q_0=0.01$ was obtained using the perturbative results in \cite{Stotyn:2011ns}, while subsequent solutions used the previously generated solution as a seed with a step size of $\Delta q_0=0.01$.  In all cases, convergence to a solution occurred after 5 iterations, where convergence was determined by the changes to the spectral coefficients being the same order of magnitude for 2 subsequent iterations; typically at convergence these changes are on the order of $10^{-30}$ or smaller.

The fact that Chebyshev approximations have exponential convergence\cite{Pfeiffer:2005zm} leads to a number of checks on the validity of the solution generated by this method.  First of all, since the Chebyshev polynomials are all of order 1, and the $j^{\mathrm{th}}$ spectral coefficient is approximately $C_j\approx C_0e^{-kj}$ for some constant $k$ that depends on the approximated function, the expansion at any point obeys the inequality
\begin{equation}
q\le\sum_{j=0}^{N}C_0 e^{-kj}\sim \int_{0}^{N}{C_0 e^{-kj}dj}.
\end{equation}
A rough estimate of the error of the approximation is then given by
\begin{equation}
\mathrm{error}\approx \int_{N}^{\infty}{C_0 e^{-kj}dj} = \frac{C_0 e^{-kN}}{k}
\end{equation}
where the term in the numerator is the $N^{\mathrm{th}}$ spectral coefficient, $C_N$.  Therefore, an order of magnitude estimate of the percent-error in the Chebyshev expansion is simply given by
\begin{equation}
\mathrm{\%~error}\approx \frac{C_N}{C_0},
\end{equation}
which is easily computable by inspection of the output coefficients of an iteration.  Indeed, ensuring this quantity maintained a value below $10^{-8}$ was one of the criteria used to determine whether the Chebyshev grid used was sufficiently dense.  

Similarly, because of the exponential convergence of the spectral coefficients, a plot of $\log|C_j|$ vs $j$ should be approximately a straight line with a negative slope $-k$.  This property was used extensively to determine whether there was a sufficient number of grid points or, equally as important, a sufficient amount of digit precision in the calculations.  Since the spectral coefficients decrease exponentially, after a certain point, $j=n$, the digit precision of the calculations, denoted Prec, cannot resolve the difference between spectral coefficients: the remaining coefficients $C_j$ for $j=n,...,N$ are then numerical noise.  To overcome this obstacle, we always worked with Prec$~\gtrsim N$ and found that plots of $\log|C_j|$ vs $j$ behaved as expected.

Unfortunately, as $q_0$ increases, this becomes computationally quite expensive and time intensive.  When $q_0$ is sufficiently small, the curvature scale is small everywhere and a sparse Chebyshev grid of $N\sim 20$ suffices to solve the equations of motion.  As $q_0$ begins to rise, both $N$ and Prec need to correspondingly increase to resolve the curvature, with the time for each iteration being at least quadratic in both.  We found that we typically reached the point of diminishing returns at approximately $N=100$, which was the upper limit used to generate the results of the following section.

Finally, we note that the above considerations are not always sufficient to determine whether the generated solutions are viable.  To determine this, we imposed that the first law of thermodynamics must be satisfied to one part in $10^6$.  We found that our code would continue to generate solutions well beyond this limit on the first law if we blindly pushed it forward.  However, if we evaluated the equations of motion for such spurious solutions at points in-between the Chebyshev grid points, we found that they were indeed not satisfied within the required tolerance, whereas they were satisfied for solutions obeying the first law bound.  This was primarily how we decided when to increase the grid density and digit precision from one solution to the next.

Representative plots of the metric functions and scalar field function constructed by the above numerical methods are plotted in figure \ref{fig:functions} in appendix A: these plots are for $D=5$ and $q_0=5.0$, although the generic features of the plots are found to change very little with varying dimension or varying $q_0$.  It is clear that all of these functions have their expected behaviours and are free from singularities, except for $f(y)$ which diverges as $y\rightarrow 1$ as one expects due to the AdS boundary conditions.

\section{Results and Discussion}
\label{Results}

With confidence in the validity of our generated solutions assured by the methods detailed above, we can now present the results.  Of course, the physically interesting data for these boson stars amount to their physical properties, all of which can be written explicitly in terms of the boundary values of the analytic $q$ functions.  For instance, the Kretschmann invariant, $K=R_{abcd}R^{abcd}$, at the centre of the boson star is given by
\begin{equation}
K_{n}=\frac{2\left(n+1\right)}{\ell^4}\left[\left(n+2\right)+\frac{3\left(n-1\right)\left(n+3\right)}{n}q_{h}'(0)^{2}-4q_{0}^{2}+\frac{4}{n}q_{0}^{4}\right],
\end{equation}
while the thermodynamic quantities take the form \cite{Ashtekar:1999jx,Das:2000cu}
\begin{align}
M={}&\frac{\pi^{\frac{n-1}{2}}\ell^{n-1}}{8\left(\frac{n-1}{2}\right)!}\big((n+1)q_h(1)-nq_f(1)\big),\\
J={}&\frac{(-1)^{\frac{n+1}{2}}\pi^{\frac{n-1}{2}}\ell^n}{4\left(\left(\frac{n-1}{2}\right)!\right)^2}q_\Omega^{(\frac{n+1}{2})}(1),\\
\omega={}&-q_\Omega(1).
\end{align}
as can be shown using equations (\ref{EJ}) and (\ref{fq} -- \ref{Pq}).
Similarly, the perturbative parameter $\epsilon$ used in \cite{Stotyn:2011ns} is given by
\begin{equation}
\epsilon=q_\Pi(1).
\end{equation}
Given these quantities, we present here the results for $n=3,5,7,9$, where we
include $n=3$ both for completeness, and to demonstrate that we reproduce the existing results
of \cite{Dias:2011at}.
\vskip 0.5cm

\begin{figure}[ht!]
     \begin{center}
        \subfigure[~$M$ vs $q_{0}$;  inset is a close-up of the maximum for $n=9$.]{%
            \label{fig:first}
            \includegraphics[width=0.475
           \textwidth]{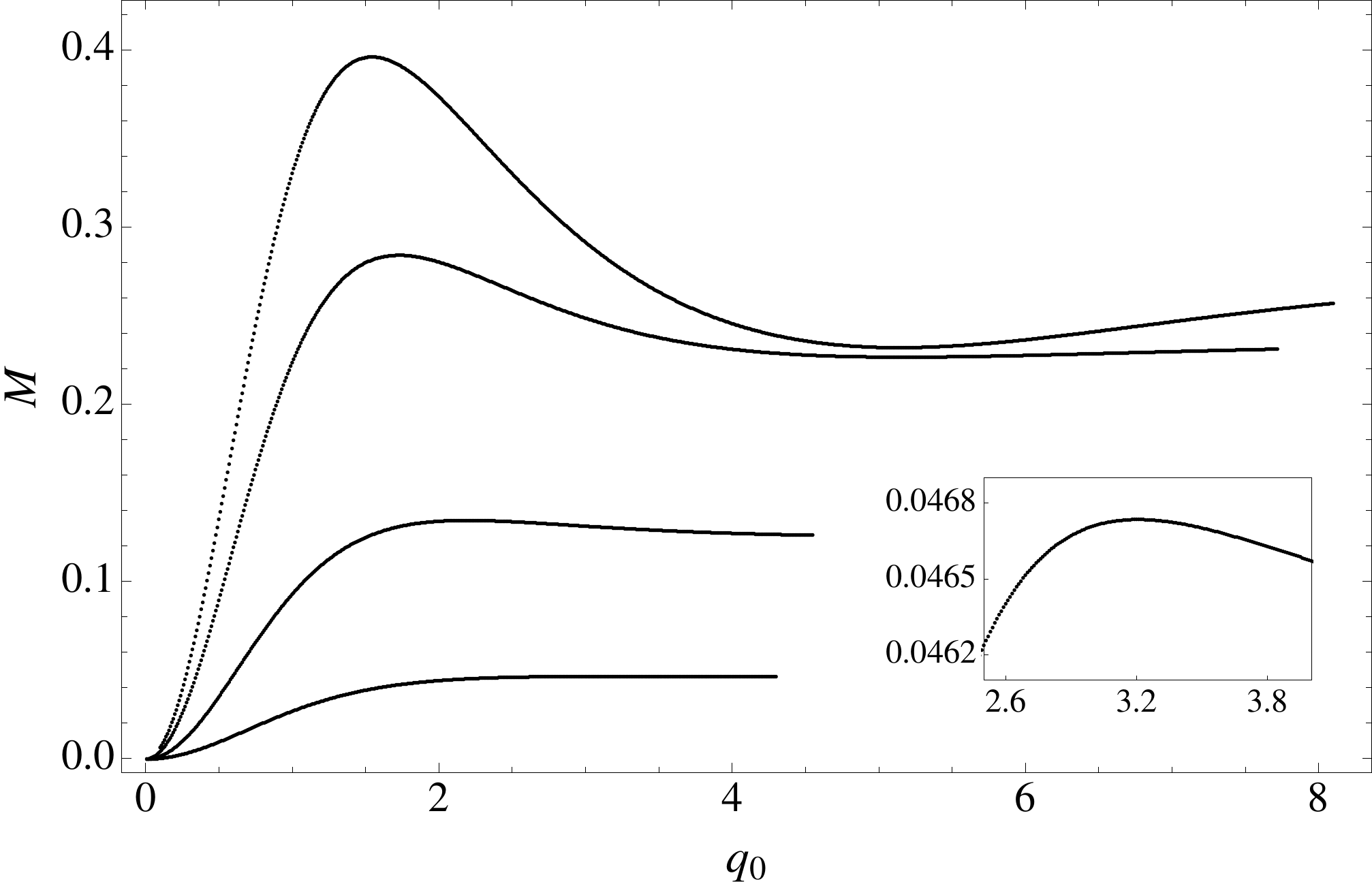}
            \put(-180,142){\tiny $n=3$}
            \put(-175,108){\tiny $n=5$}
            \put(-160,64){\tiny $n=7$}
            \put(-135,37){\tiny $n=9$}
            \put(-50,64){ \tiny $n=9$}
        }%
        \subfigure[~$J$ vs $q_{0}$;  inset is a close-up of the maximum for $n=9$.]{%
           \label{fig:second}
           \includegraphics[width=0.485
           \textwidth]{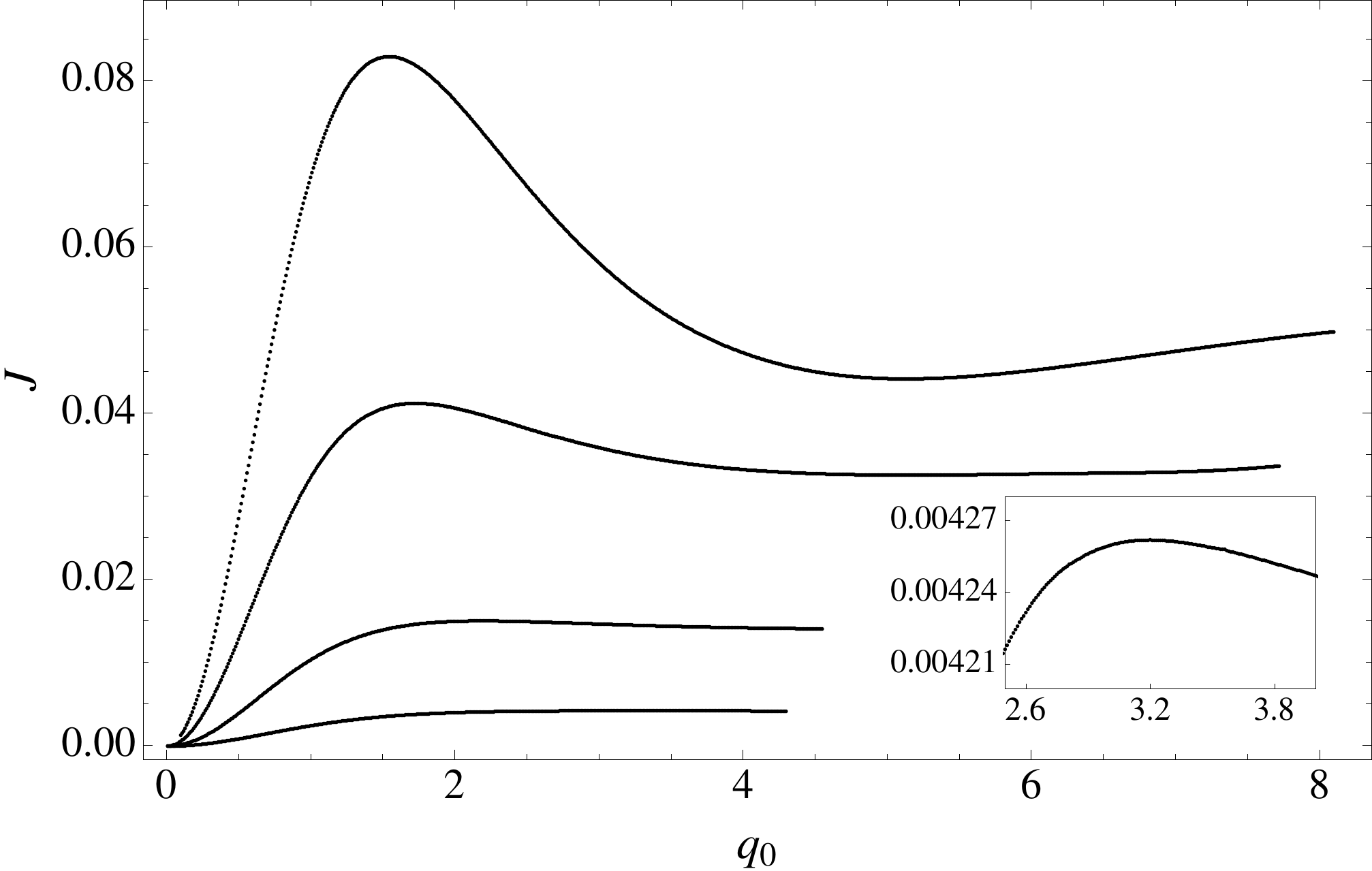}
           \put(-180,142){\tiny $n=3$}
            \put(-175,83){\tiny $n=5$}
            \put(-160,45){\tiny $n=7$}
            \put(-135,30){\tiny $n=9$}
            \put(-50,60){ \tiny $n=9$}
        }\\ %  ------- End of the first row ----------------------%
        \subfigure[~$\omega$ vs $q_{0}$;  inset is a close-up of the minima for $n=7,9$.]{%
            \label{fig:third}
            \includegraphics[width=0.48\textwidth]{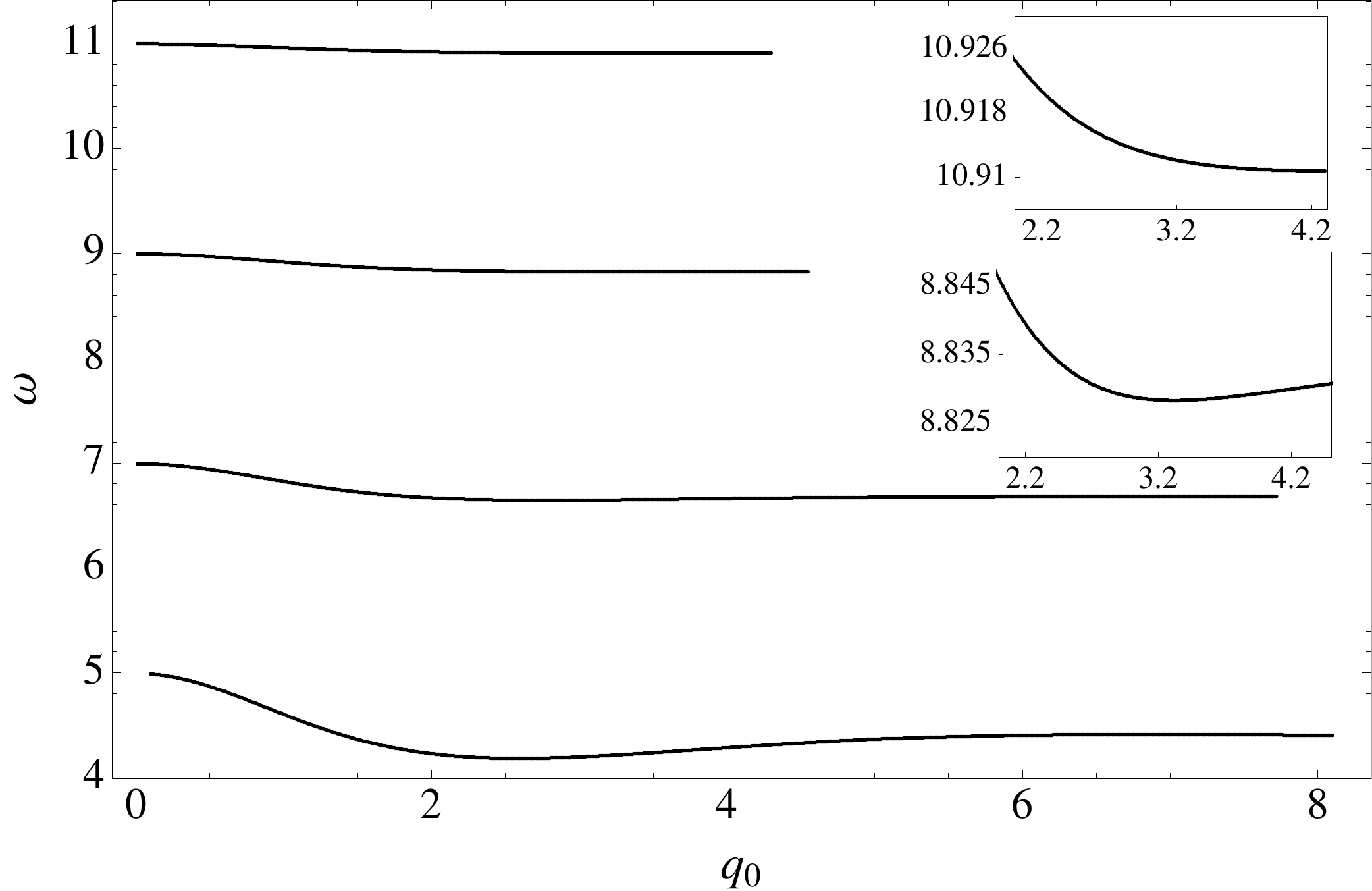}
            \put(-45,144){\tiny $n=9$}
            \put(-45,104){\tiny $n=7$}
            \put(-160,145){\tiny $n=9$}
            \put(-160,107){\tiny $n=7$}
            \put(-160,69){\tiny $n=5$}
            \put(-160,25){\tiny $n=3$}
        }%
        \subfigure[~$\epsilon$ vs $q_{0}$;  inset is a close-up of the maximum for $n=9$.]{%
            \label{fig:fourth}
            \includegraphics[width=0.48\textwidth]{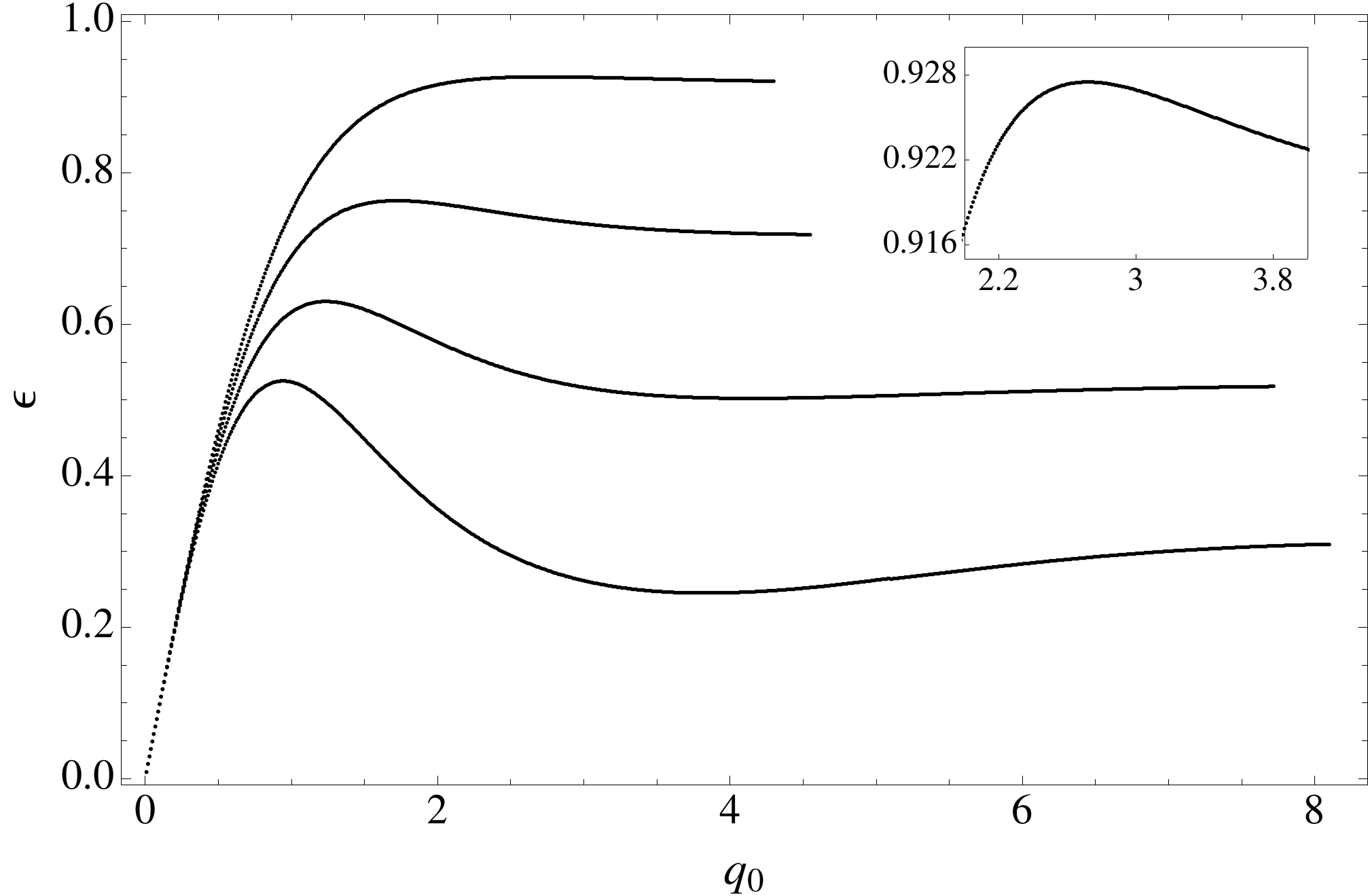}
            \put(-127,142){\tiny $n=9$}
            \put(-127,116){\tiny $n=7$}
            \put(-127,54){\tiny $n=3$}
            \put(-127,88){\tiny $n=5$}
            \put(-52,141.5){ \tiny $n=9$}
        }%
    \end{center}
    \caption{The boson star mass $(a)$, angular momentum $(b)$, angular frequency $(c)$, and perturbative parameter $\epsilon$ $(d)$ plotted against the parameter $q_0=\ell\Pi'(0)$.  In each dimension, these quantities all display damped harmonic oscillations and asymptote to finite values as $q_0\rightarrow\infty$.
     }%
   \label{fig:subfigures}
\end{figure}
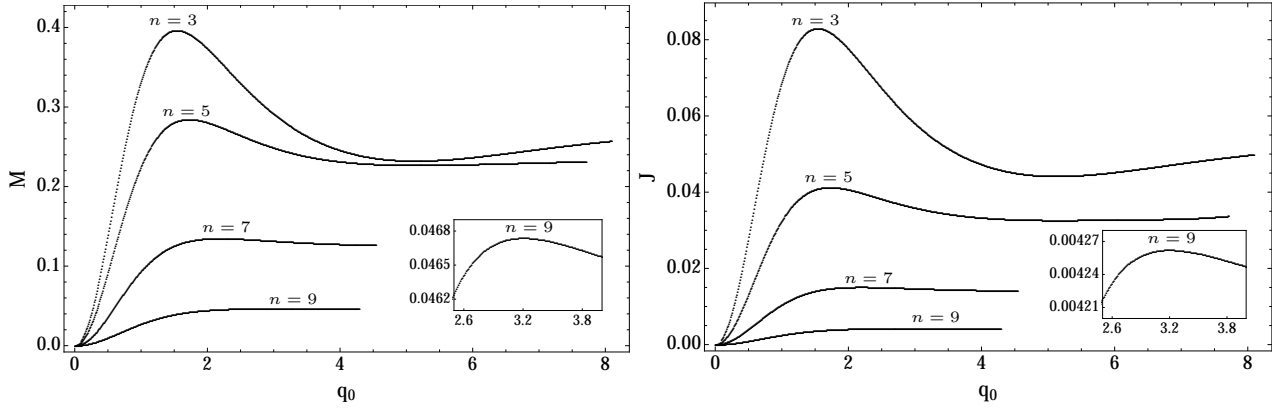
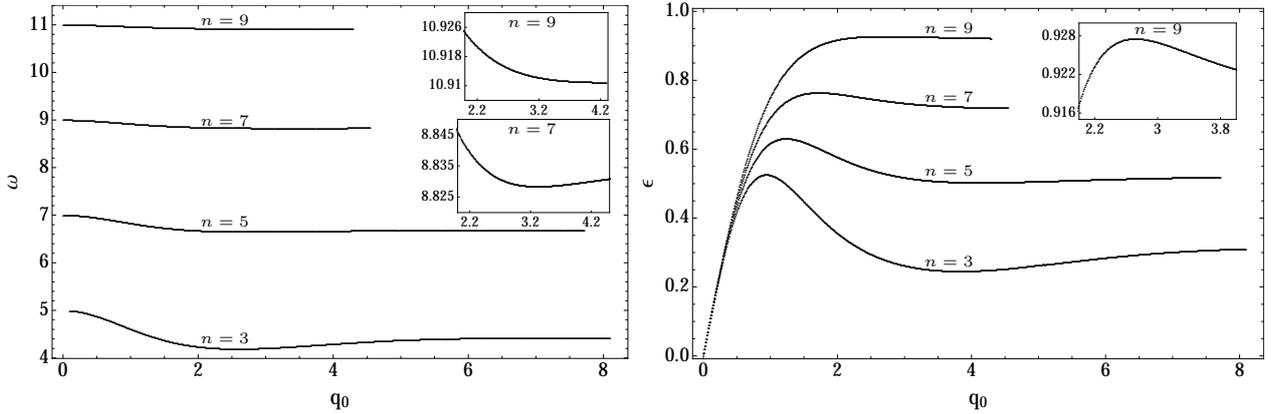

 In all dimensions we find behaviour that is qualitatively similar to that
which was found in \cite{Dias:2011at} for $n=3$.  In figure \ref{fig:subfigures}, we plot mass, $M$, angular momentum, $J$, angular velocity, $\omega$, and asymptotic scalar field amplitude, $\epsilon$, as functions of $q_{0}$, which can be seen to uniquely parameterize the family of boson stars in each dimension.

We find numerically that each of these quantities displays damped harmonic oscillations as $q_0$ increases, with the amplitude of the oscillation decreasing as the dimension increases.  Furthermore, as $q_0$ tends toward infinity, these quantities all asymptote to finite central values that depend on the dimension.  

The damped oscillations have interesting consequences.  For instance, the maxima of the $M$ vs $q_0$ graph represent instabilities in the boson stars: adding a small but finite perturbation with energy $\delta M$ will necessarily force the solution off the family of boson stars, likely forming a (potentially hairy) black hole as a result.  It is tantalizing to conjecture that this effect may be intimately tied to the turbulent instability in AdS spaces.  Recall that AdS is non-perturbatively unstable to forming a black hole because of the large number of resonant frequencies coming from the normal modes being integer multiples of the AdS frequency\cite{Dias:2012tq}.  Perhaps similar normal mode resonances are occurring at the boson star mass maxima, leading to non-perturbative instabilities toward black hole formation at these particular values of $q_0$.  From the holographic dual point of view, this would imply that generally the strongly coupled dual CFT states would not thermalize, except at particular points in parameter space corresponding to an infinite tower of resonances.

The interpretation of the mass minima is less straightforward, although these points are perturbatively unstable in parameter space in the sense that a small perturbation could push the solution either to the left or the right along the family of boson stars.  The holographic dual interpretation is similarly obscured due to a lack of a map between the gravitational degrees of freedom and those in the CFT for these objects.  Nevertheless, these minimum mass states still represent CFT states that never thermalize since small but finite perturbations keep the solution within the family of boson stars.  That being said, from the perturbative results of \cite{Stotyn:2011ns} we know that in the perturbative regime (\emph{i.e.} in the vicinity of the first minimum located at $q_0=0$) the boson stars are perturbatively joined to small hairy black holes.  This is presumably because the geometry of small boson stars is still sufficiently close to AdS that approximate resonances still exist in the normal modes, causing the instability.  It is far from obvious whether the subsequent mass minima have similar behaviour, like the mass maxima do.  We leave this, and other issues of stability, as an open question for future consideration.

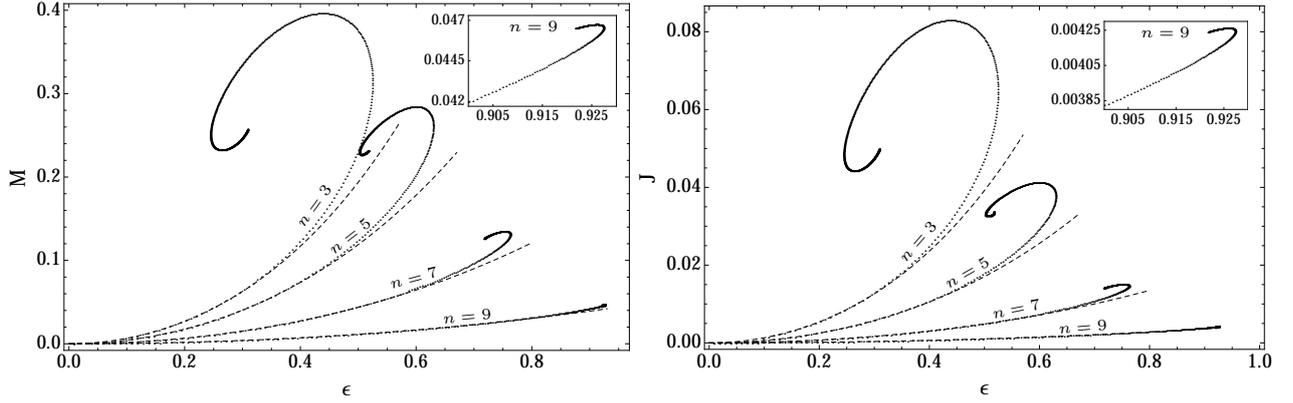
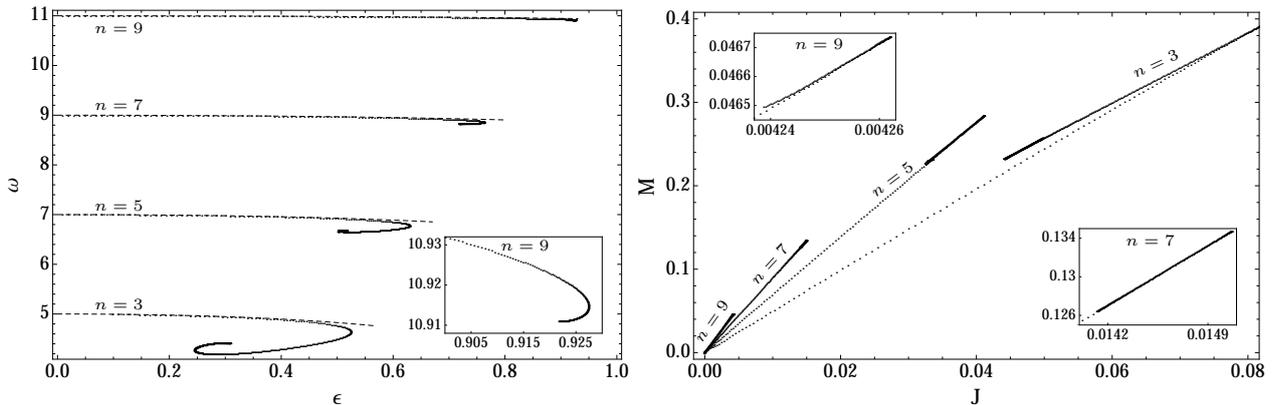
\begin{figure}[ht!]
     \begin{center}
        \subfigure[~$M$ vs $\epsilon$;  inset is a close-up of the end of the $n=9$ line.]{%
            \label{fig:2first}
            \includegraphics[width=0.475
           \textwidth]{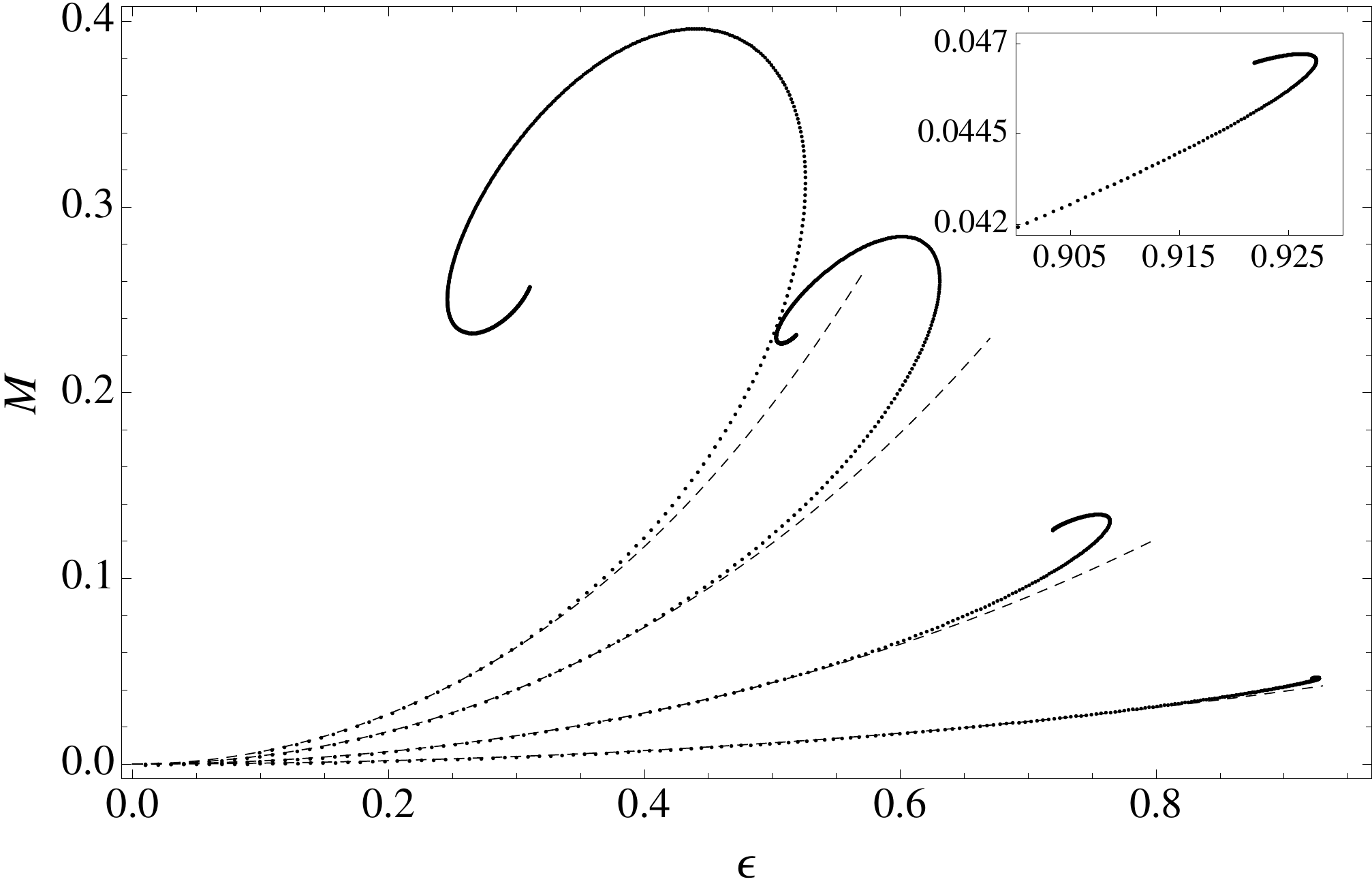}
            \put(-45,139){\tiny $n=9$}
             \put(-125,65){\begin{turn}{54} \tiny $n=3$ \end{turn}}
              \put(-113,55){\begin{turn}{38} \tiny $n=5$ \end{turn}}
               \put(-90,42){\begin{turn}{15} \tiny $n=7$ \end{turn}}
                \put(-70,29){\begin{turn}{5} \tiny $n=9$ \end{turn}}
        }%
        \subfigure[~$J$ vs $\epsilon$;  inset is a close-up of the end of the $n=9$ line.]{%
           \label{fig:2second}
           \includegraphics[width=0.485
           \textwidth]{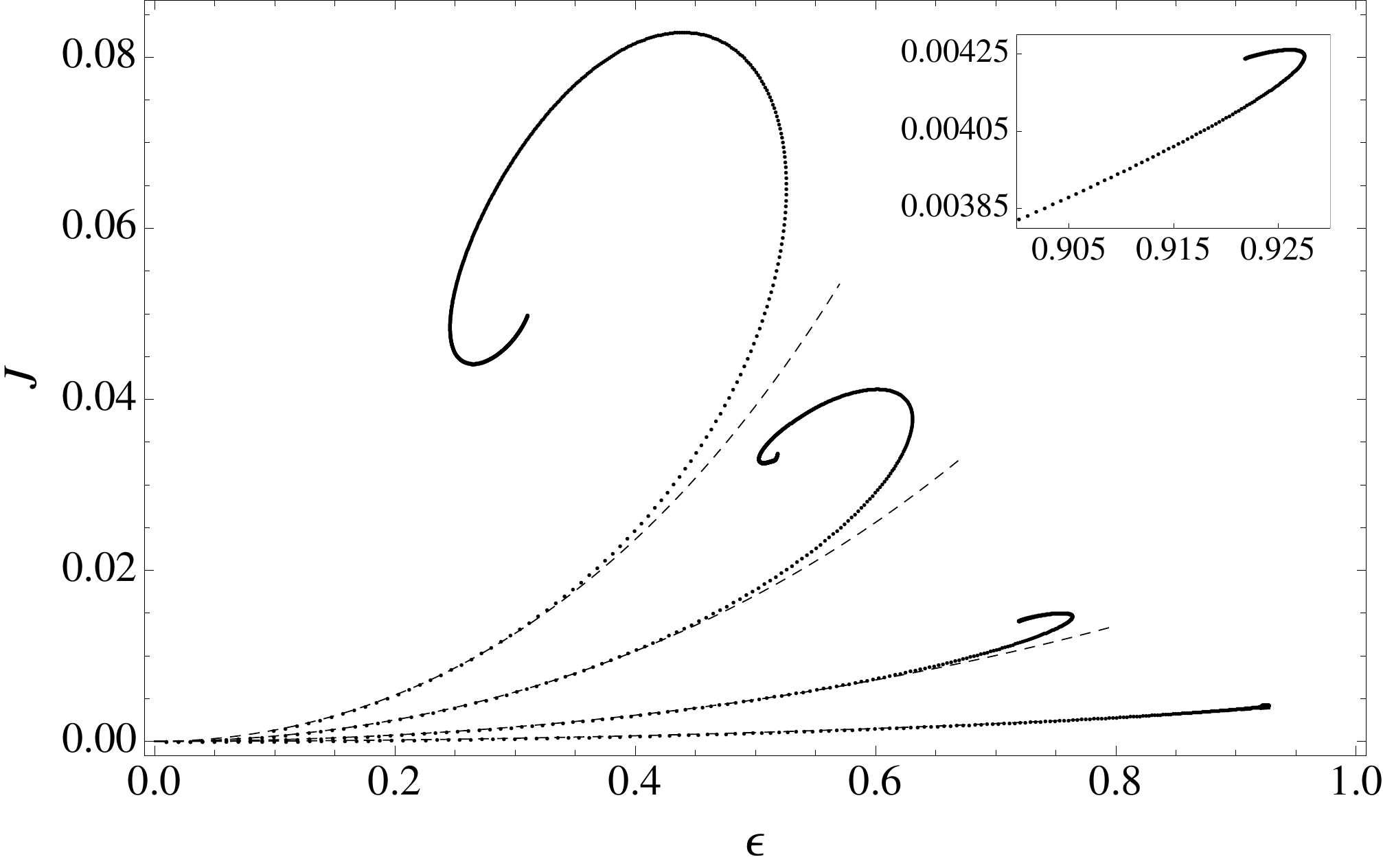}
            \put(-49,137){\tiny $n=9$}
            \put(-140,52){\begin{turn}{49} \tiny $n=3$ \end{turn}}
              \put(-123,42){\begin{turn}{30} \tiny $n=5$ \end{turn}}
               \put(-105,31){\begin{turn}{11} \tiny $n=7$ \end{turn}}
                \put(-80,25){\begin{turn}{5} \tiny $n=9$ \end{turn}}
        }\\ %  ------- End of the first row ----------------------%
        \subfigure[~$\omega$ vs $\epsilon$;  inset is a close-up of the end of the $n=9$ line.]{%
            \label{fig:2third}
            \includegraphics[width=0.48\textwidth]{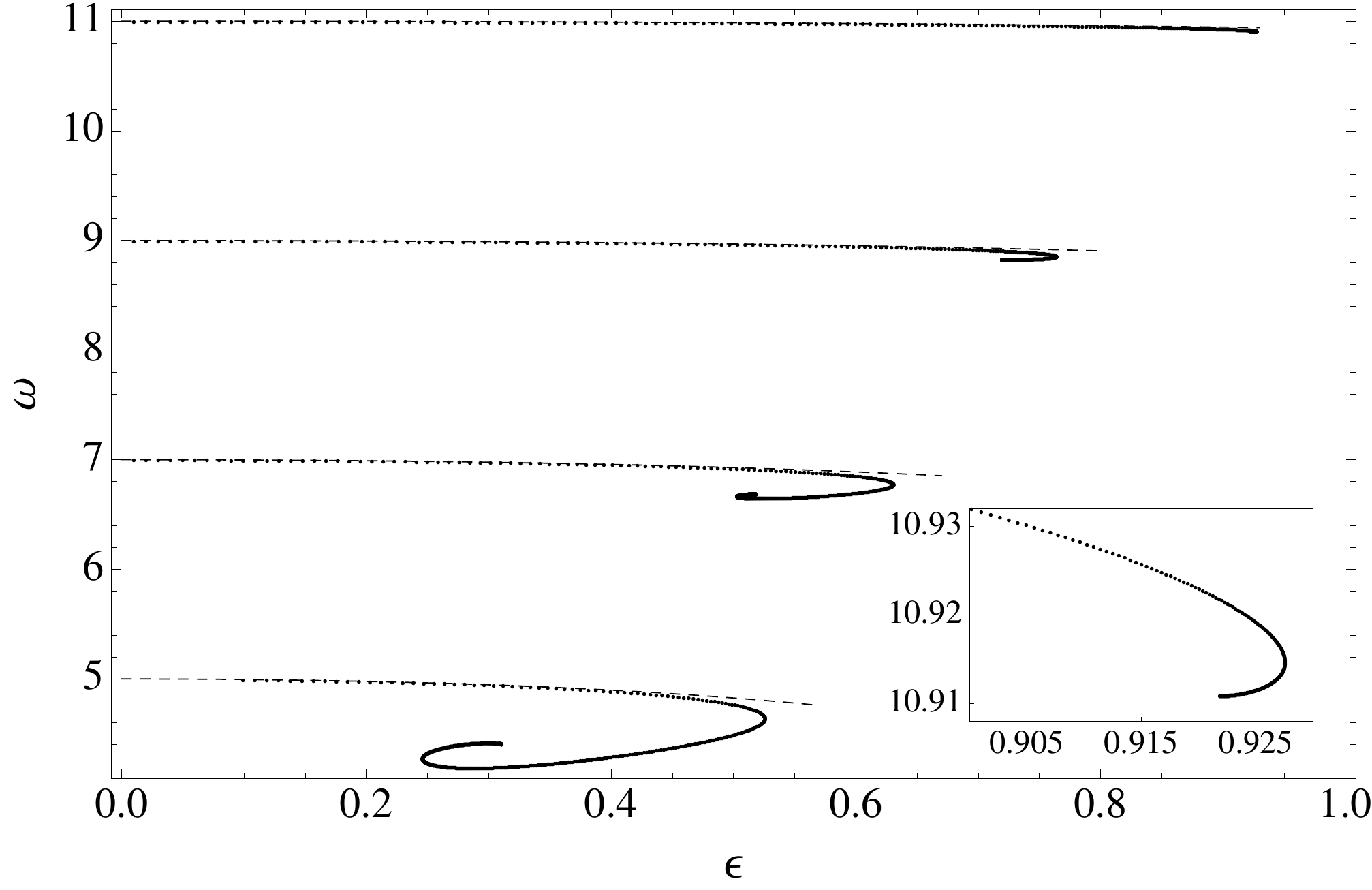}
            \put(-48,60){\tiny $n=9$}
            \put(-200,142){\tiny $n=9$}
            \put(-200,113){\tiny $n=7$}
            \put(-200,75){\tiny $n=5$}
            \put(-200,38){\tiny $n=3$}
        }%
        \subfigure[~$M$ vs $J$;  inset is a close-up of the end of the $n=7,9$ lines.]{%
            \label{fig:2fourth}
            \includegraphics[width=0.48\textwidth]{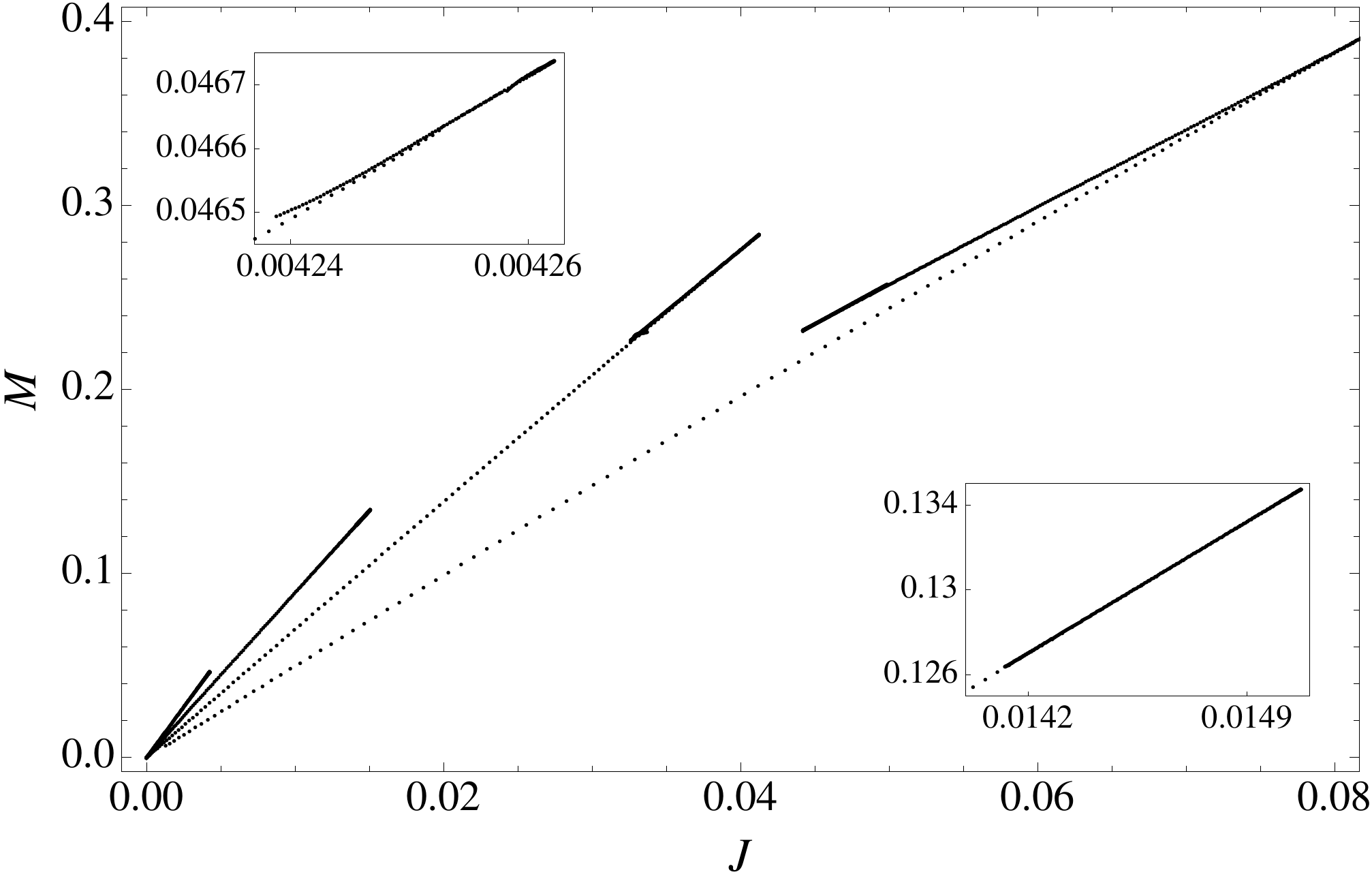}
            \put(-52,62){\tiny $n=7$}
            \put(-176,136){\tiny $n=9$}
            \put(-214,25){\begin{turn}{55} \tiny $n=9$ \end{turn}}
            \put(-193,47){\begin{turn}{48} \tiny $n=7$ \end{turn}}
            \put(-148,80){\begin{turn}{37} \tiny $n=5$ \end{turn}}
            \put(-50,125){\begin{turn}{25} \tiny $n=3$ \end{turn}}
        }%
    \end{center}
    \caption{%
        The boson star mass $(a)$, angular momentum $(b)$, and angular frequency $(c)$ plotted against the perturbative parameter $\epsilon$ with the perturbative results from \cite{Stotyn:2011ns} plotted as dashed lines.  In the non-perturbative regime, $\epsilon$ no longer uniquely parameterizes the boson star family.  The spiral behavior of $(a),~(b),~\mathrm{and}~(c)$ manifests as a jagged zig-zag in plot $(d)$ of $M$ vs $J$.  This is a consequence of the first law.  Note that in plot $(d)$ the perturbative results of \cite{Stotyn:2011ns} are not included; each point is a solution at a particular value of $q_0$ and these points become increasingly dense as $q_0$ increases.
     }%
   \label{fig:subfigures2}
\end{figure}

In the perturbative regime of Ref. \cite{Stotyn:2011ns}, the parameter $\epsilon$ uniquely determined the boson star solutions.  In the full numerical treatment this is no longer the case, as can be seen in figure \ref{fig:subfigures2}.  The damped oscillations with respect to $q_0$ seen in figure \ref{fig:subfigures} show up as damped spirals with respect to $\epsilon$ in figure \ref{fig:subfigures2}, with the centre of the spirals corresponding to the asymptotic values attained as $q_0\rightarrow\infty$.  Furthermore, plots of the Kretschmann scalar, $K$, against $\epsilon$ shows that the curvature at the centre of the boson star is quickly diverging as $q_0$ increases.  In the limit $q_0\rightarrow\infty$, the boson star develops a curvature singularity and hence collapses to form a black hole.  We note, however, that as $q_0$ increases, the family of solutions is spiralling tighter and tighter around the asymptotic black hole values and the boson star will become non-perturbatively unstable to forming a black hole before the limit $q_0\rightarrow\infty$ is reached.  This can also be seen in figure \ref{fig:subfigures} by the fact that the $M$ vs $q_0$ graphs asymptote to horizontal lines in all dimensions, meaning that adding a small but finite perturbation will necessarily force the solution off the boson star branch, leading to the formation of a black hole.

Finally, we note in figure \ref{fig:2fourth} that plots of $M$ vs $J$ form jagged zig-zags in each dimension.  This is a direct consequence of the first law of thermodynamics, combined with the boson stars being a one parameter family of solutions.  Rewriting the first law as
\begin{equation}
\frac{dM}{dq_0}=\omega\frac{dJ}{dq_0} \label{eq:FirstLaw2}
\end{equation}
demonstrates that the extrema of the $M$ vs $q_0$ and the $J$ vs $q_0$ graphs of figure \ref{fig:subfigures} must occur at the same value of $q_0$ in a given dimension.  Such points represent the cusps of figure \ref{fig:2fourth}, while the slopes of the zigs and zags represent the value of $\omega$ by virtue of (\ref{eq:FirstLaw2}).  As the space-time dimension increases, the amplitude of oscillation for $\omega$ decreases, meaning that the zig-zags of figure \ref{fig:2fourth} become more and more collinear as can be seen by comparing $n=3,5,7,9$.  Lastly, since the amplitudes of oscillation for $M$ and $J$ decrease with increasing $q_0$, the amplitude of the $M$ vs $J$ zig-zags decrease, as can be clearly seen for $n=3,5$ in figure \ref{fig:2fourth}.

\section{Conclusions}

\label{Conclusion}

Using a relaxation procedure on a Chebyshev grid to solve the equations of motion, we have numerically obtained asymptotically AdS boson star solutions with only one Killing field in all odd space-time dimensions of interest
in string theory, \emph{i.e.} up to $D=11$.\footnote{The full numerical boson star solution in $D=3$ can be found separately in Ref. \cite{Stotyn:2013spa} due to its significantly different behaviour than its higher dimensional counterparts presented here.} Previously, these solutions had only been produced perturbatively in $D=3,5,7,9,11$, and numerically in $D=5$.  Our results reveal that the behaviour of boson stars in 7, 9 and 11 dimensions is qualitatively very similar to their 5-dimensional counterparts
originally found in \cite{Dias:2011at}. In particular, their various properties ($M$, $J$, $\omega$, and $\epsilon$) undergo damped harmonic oscillations about finite critical values as the central energy density is increased. Accordingly, we see that the amplitude of these oscilations is lower in higher space-time dimension.

Since the publication of Ref. \cite{Dias:2011at} a number of developments in the non-linear (in)stability of asymptotically AdS space-times have been made.  For instance, it was argued in \cite{Dias:2011ss} that global AdS is non-perturbatively unstable to the formation of a black hole; the heuristic idea is that the reflecting boundary conditions of AdS means that eventually the finite energy perturbations will come together in such a way that the energy is sufficiently concentrated to form a black hole.  In light of this heuristic picture, it seems plausible that \emph{any} asymptotically AdS space-time will be non-perturbatively unstable to the formation of a black hole.  However, it was then discovered in \cite{Dias:2012tq} that this instability is really due to the high level of symmetry present in global AdS, since the normal mode frequencies are all resonant with the AdS frequency.  This was vindicated in Ref. \cite{Buchel:2013uba}, which found a large class of boson star initial data that were non-perturbatively stable.

The current work mirrors the 5 dimensional solution found in \cite{Dias:2011at} and given the above developments, we are able to further elucidate the rather rich physics of these boson star solutions.  For instance, the damped oscillations in the mass implies that the boson stars corresponding to the mass maxima are all non-perturbatively unstable to black hole formation.  This could be related to the resonances of global AdS, although this remains conjecture at this point.  From the holographic dual perspective, this implies that the boson stars correspond to strongly coupled CFT states that never thermalize, except at the infinite tower of resonances corresponding to the boson star mass maxima.  Little else is known about the holographic dual description, other than that the CFT state is a localized scalar operator defining a current that breaks rotational symmetry.  Whether there are any physically realizable quantum systems that share some of these properties remains to be seen, although such systems would be very intriguing indeed, on account of the fact that they never thermalize in finite time (modulo the aforementioned exceptions).

\section*{Acknowledgements}

This work was supported in part by the Natural Sciences and Engineering
Research Council of Canada. D.L. was funded in part by the Rhodes Trust. M.O. acknowledges additional support from the University of Waterloo Department of Applied Mathematics. We would like to thank M.Chanona, H. Pfeiffer, O. Dias, and J. Santos 
for helpful discussions at various stages of this project.

\newpage

\appendix

\section{Representative Plots of the Metric and Scalar Field Functions}

\begin{figure}[ht!]
     \begin{center}
        \subfigure[~Metric function $f(y)$ for $D=5$ and $q_0=5.0$]{%
            \label{fig:f}
            \includegraphics[width=0.475 \textwidth]{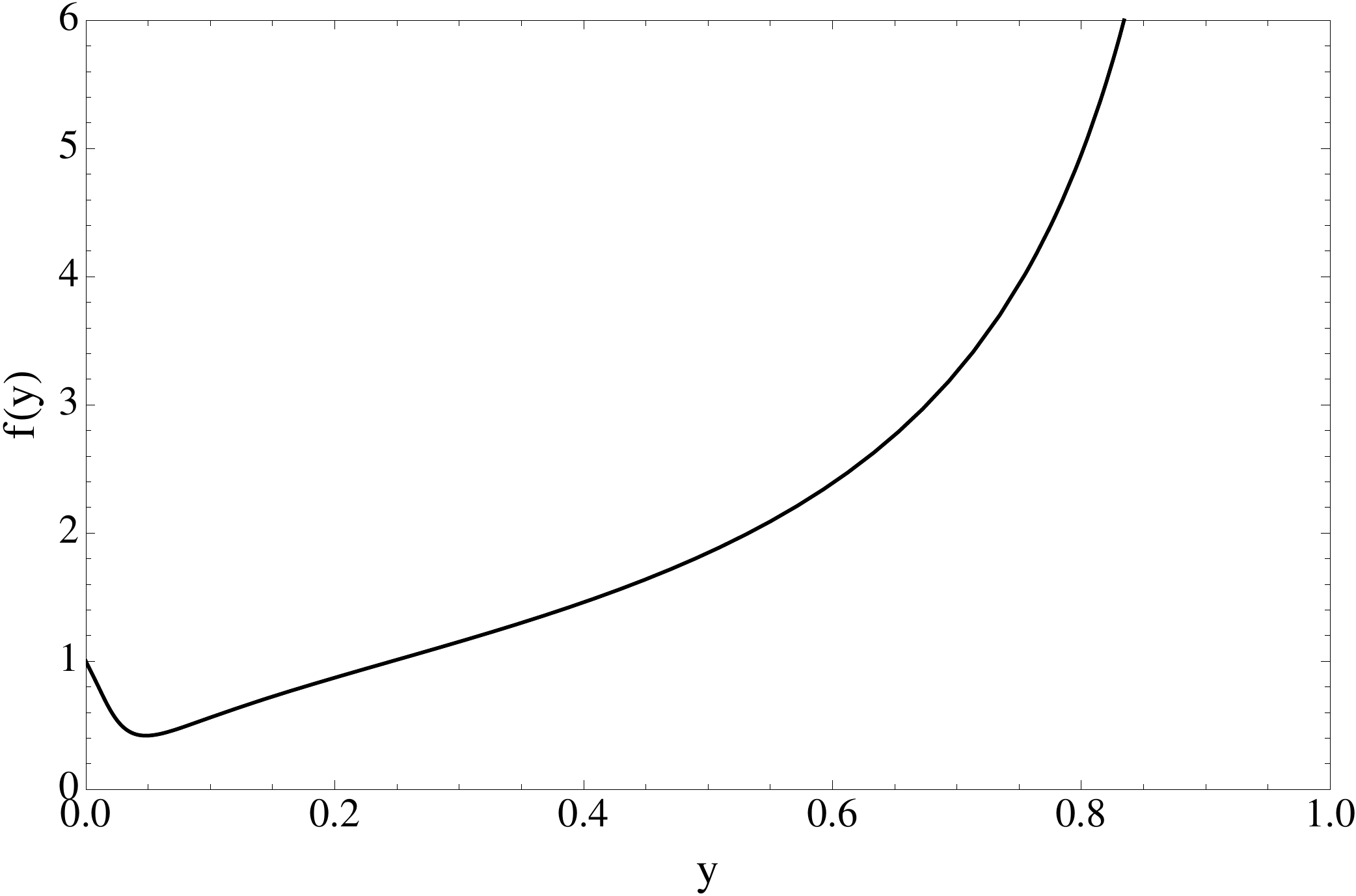}
        }%
        \subfigure[~Metric function $g(y)$ for $D=5$ and $q_0=5.0$]{%
            \label{fig:g}
            \includegraphics[width=0.475
           \textwidth]{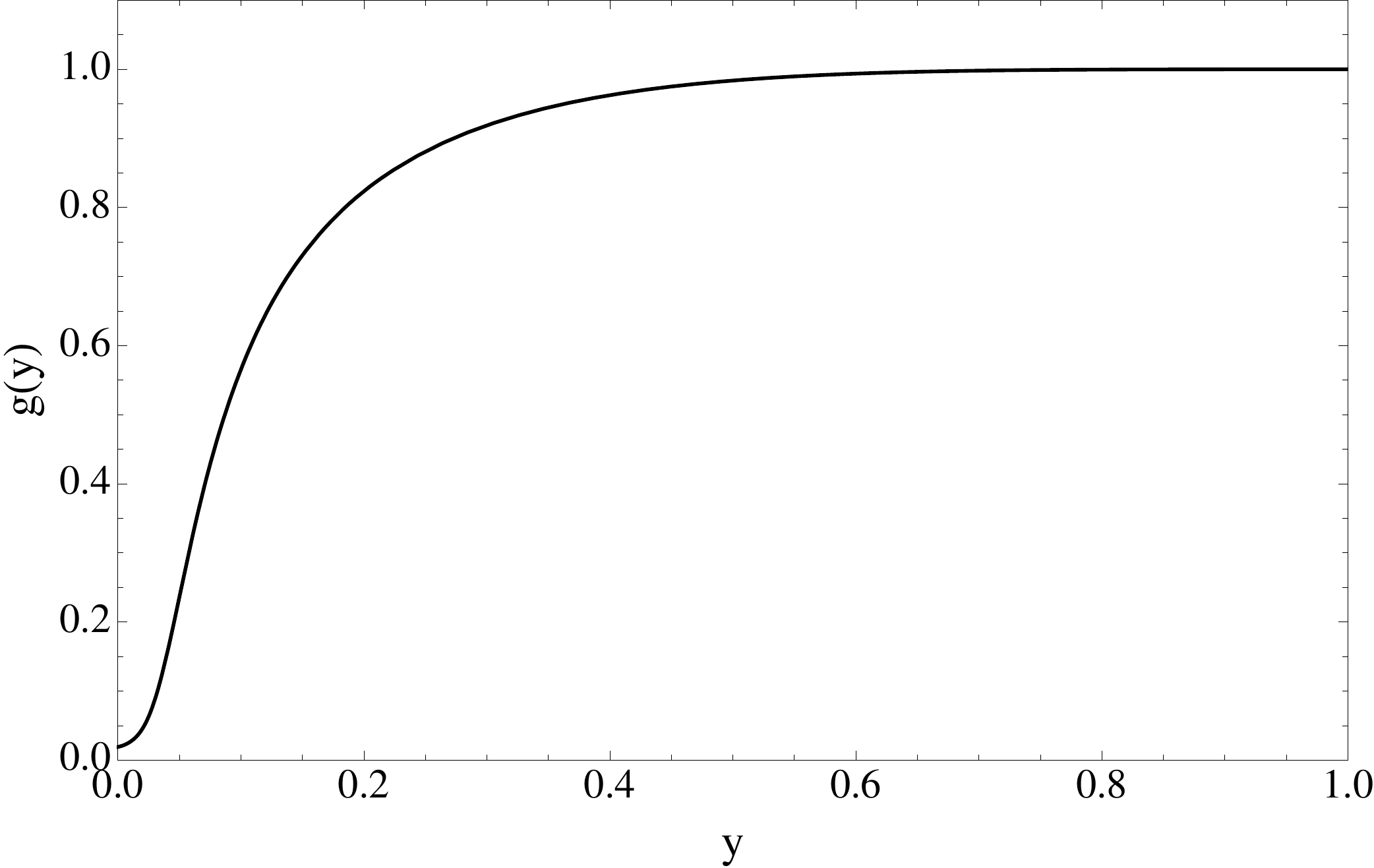}
        }\\ 
       %  ------- End of the first row ----------------------%
        \subfigure[~Metric function $h(y)$ for $D=5$ and $q_0=5.0$]{%
            \label{fig:h}
            \includegraphics[width=0.475
           \textwidth]{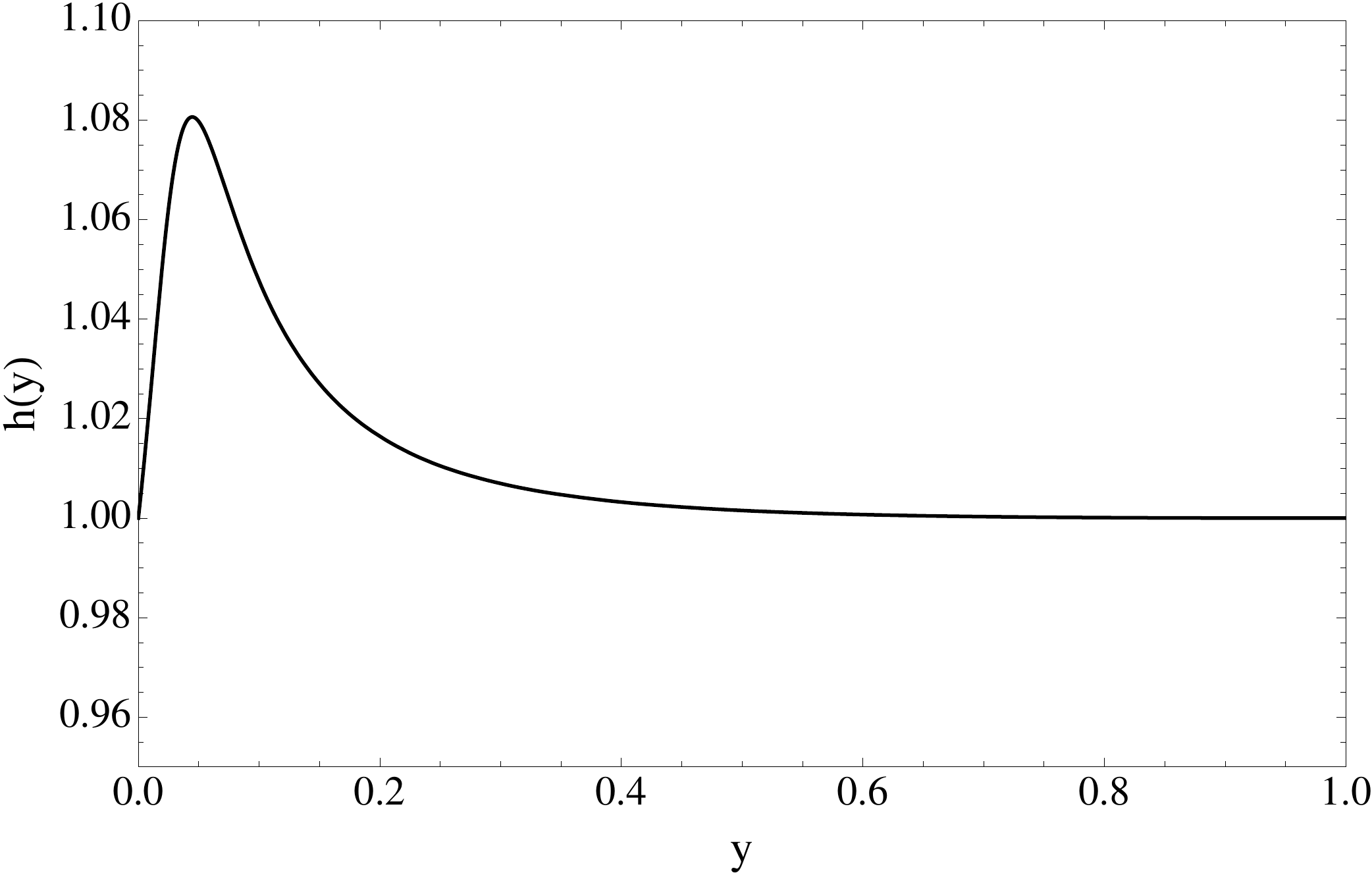}
        }%
        \subfigure[~Metric function $\Omega(y)$ for $D=5$ and $q_0=5.0$]{%
            \label{fig:Omega}
            \includegraphics[width=0.475
           \textwidth]{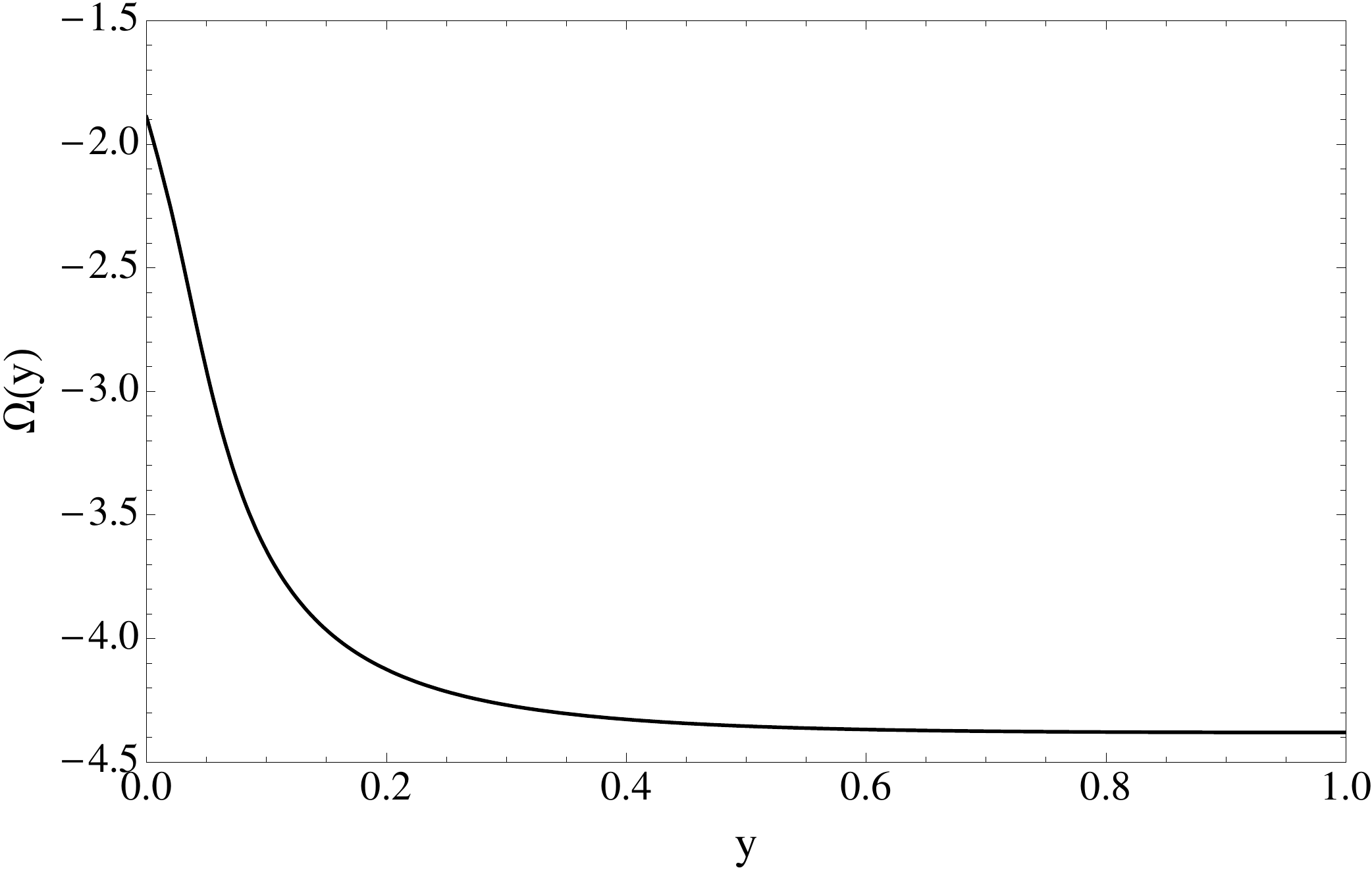}
        }\\ %  ------- End of the second row ----------------------%
        \subfigure[~Metric function $\Pi(y)$ for $D=5$ and $q_0=5.0$]{%
            \label{fig:Pi}
            \includegraphics[width=0.475
           \textwidth]{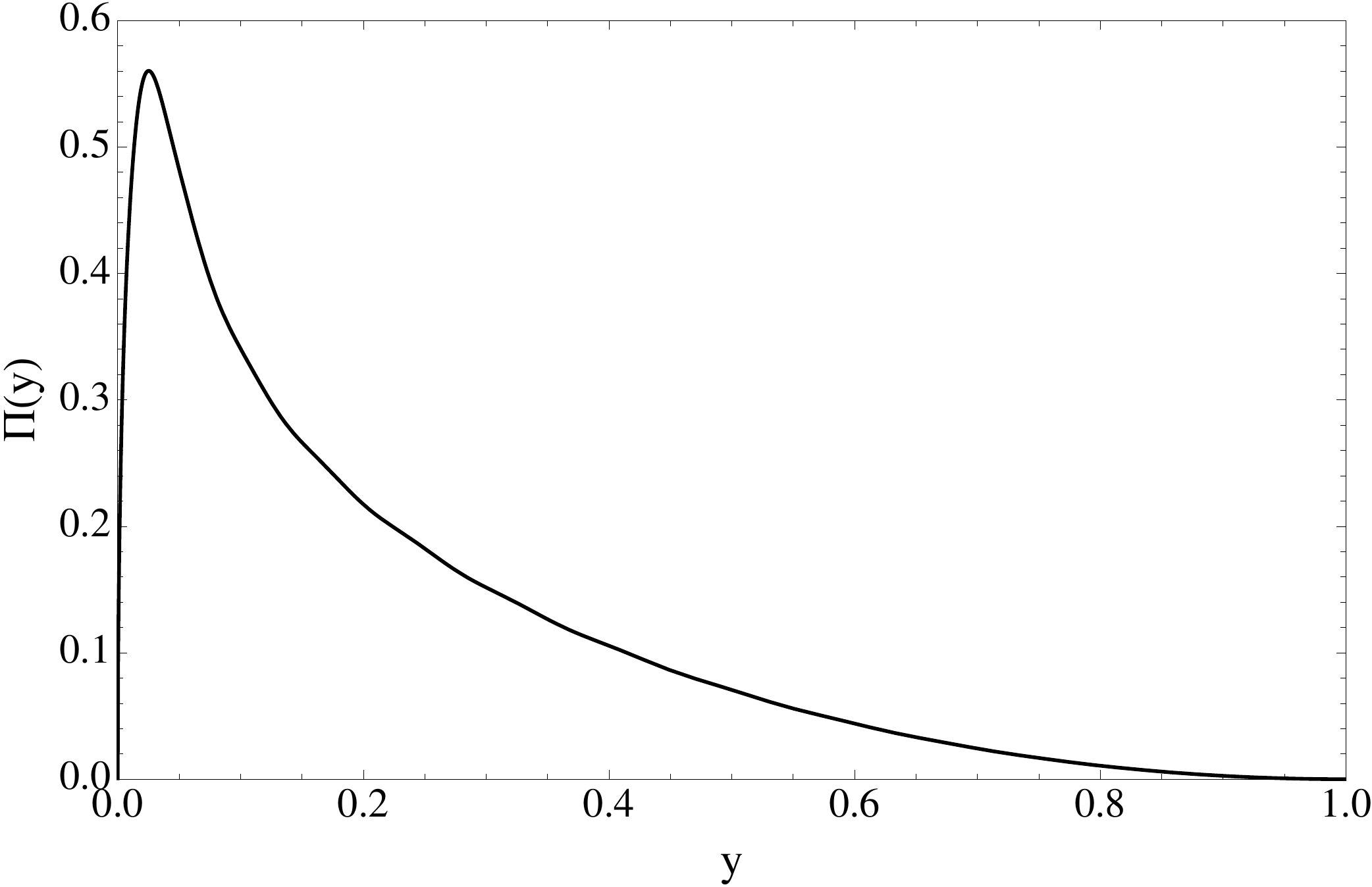}
        }
    \end{center}
     \caption{Plots of the metric functions, $f(y),~g(y),~h(y)$, and $\Omega(y)$, and the scalar field function, $\Pi(y)$, for a value of $q_0=5.0$ in $D=5$.  We explicitly checked other values of $q_0$ for $D=5,7,9,11$ and found that while the vertical axes may have different ranges depending on the dimension and choice of $q_0$, the qualitative features of the graphs remain the same.
     }%
   \label{fig:functions}
\end{figure}

\bibliographystyle{plain}

\end{document}